\documentstyle[sprocl]{article}

\input{epsf}

\def\Journal#1#2#3#4{{#1} {\bf #2}, #3 (#4)}

\def\PRL{\em Phys. Rev. Lett.}
\def\PRB{{\em Phys. Rev.} B}

\def\be{\begin{equation}}
\def\ee{\end{equation}}
\def\bea{\begin{eqnarray}}
\def\eea{\end{eqnarray}}

\def\vk{{\bf k}}

\begin{document}

\title{PATH INTEGRAL FOR COMPOSITE FERMIONS\\ IN THE HALF-FILLED LOWEST LANDAU LEVEL}

\author{W. WELLER, J. DIETEL, TH. KOSCHNY}

\address{Institut f{\"u}r Theoretische Physik, Universit{\"a}t Leipzig,
Augustusplatz 10,\\ 04109 Leipzig, Germany\\
E-mail: wolfgang.weller@itp.uni-leipzig.de}

\author{W. APEL}

\address{Physikalisch-Technische Bundesanstalt, Bundesallee 100,\\
38116 Braunschweig, Germany}

\maketitle\abstracts{
We consider electrons in two dimensions in a strong magnetic field at
half filling of the
lowest Landau level using the Chern-Simons approach.
Starting from a lattice Hamiltonian for the electrons, we derive
a path integral (PI) formulation for the composite fermions (CF) which
respects the order of the operators.
We use a time lattice with intermediate times in order to have a PI
expressed in density fluctuations. This formulation reveals that there is no
infrared (IR) singularity in the grand-canonical potential in lowest
order perturbation theory.}

\section{Introduction}
The Chern-Simons (CS) approach to two-dimensional electrons in a strong
perpendicular magnetic field at (or close to) half filling of the lowest
Landau level, as pioneered by Halperin, Lee and Read~\cite{HLR},
is well supported by experiments. The interpretation of these
 experiments suggests
that at half filling, the composite fermions,
each consisting of an electron with two attached flux quanta, form a
Fermi liquid in zero magnetic field. Theoretically, this
approach suffers from IR singularities appearing as $\ln F$
and $\ln|\omega|$ (here $F$ is the area of the square
sample and $\omega$ the
frequency of, for example, the self energy). In this contribution, we
want to reconsider
the $\ln F$-singularities.

Recently, we studied the grand-canonical potential~\cite{DKAW} of
 the CF in the random-phase approximation
(RPA) using the CF interaction in the form
\bea
\label{H}
H^{int} =  \frac{1}{2F}  \sum_{\vk_1, \; \vk_2, \; \vk (\neq 0)} \;
     W_{ \vk ;\, \vk_1, \vk_2 } \;
     \hat{\psi}^{\dagger}_{\vk_1} \hat{\psi}^{\dagger}_{\vk_2 -\vk}
     \hat{\psi}^{}_{\vk_2} \hat{\psi}^{}_{\vk_1 - \vk}  \;\;, \\
\label{W}
W_{ \vk ;\, \vk_1, \vk_2 } =
\frac{(2\pi \hbar \tilde \phi)^2}{m_b} \, \frac{N}{F} \, \frac{1}{k^2}
  + \frac{4\pi \hbar^2 \tilde \phi}{m_b} \; i \left( \vk_1
     \times \frac{\vk}{k^2} \right) {\bf e}_z
  + \frac{2\pi \epsilon^2}{k}     \; . \nonumber
\eea
Here, $\hat{\psi}_{\vk}$
annihilates a (CF) with wave vector $\vk$,
the filling factor is $1/\tilde \phi = 1/2$,
$m_b$ is the electron band mass.
The interaction (\ref{H}) corresponds to the standard PI formulation in the
time continuum.
In the interaction $W$, the first term results from the square of the CS vector
potential \mbox{\boldmath${\cal A}$}, the second from
the linear coupling to
\mbox{\boldmath${\cal A}$} (${\bf e}_z$ is the unit vector in $z$ direction,
orthogonal to the system),
and the last term is the Coulomb interaction
($\epsilon^2 = e^2/(4\pi \varepsilon_r \varepsilon_0)$,
$\varepsilon_r$ is the dielectric constant).
In formulating the electron Hamiltonian in terms of the CF operators,
the usual approximation~\cite{HLR} replaces in
 the \mbox{{\boldmath${\cal A}$}$^2$}-term
the external pair of operators, $\hat{\psi}^{\dagger} \ldots \hat{\psi}$,
by the average electron density $\rho _{el} =N/F$
($N$ is the mean electron number).
That led to the first term in $W$.

The random-phase approximation~\cite{DKAW} gave the following main results: \\
(1) The first order (exchange) diagram for the grand-canonical potential $\Omega$
diverges in the RPA as
$\; -\beta\Omega_{\mbox{\scriptsize ex}}
    \sim  (N  \beta \mu {\tilde \phi}^2/2)  \;
           \ln \, (k_F/k_{min}) + \mbox{const}$,
where $k_{min} \propto 1/\sqrt{F}$,
$k_F$ is the Fermi momentum, and
$\mu$ the chemical potential. \\
(2) The first-order contribution of the Coulomb interaction to the
ground state energy is
$E^{(1)}_{\mbox{\scriptsize RPA}}/N = -0.6\; \epsilon^2/l_B$
($l_B$ is the magnetic length).
This compares reasonably well with
$E^{(1)}_{\mbox{\scriptsize sim}}/N = -0.466\; \epsilon^2/l_B$
obtained in numerical simulations in the spherical geometry
by Morf and d'Ambrumenil~\cite{MA}. \\
(3) The number of magnetoplasmon oscillators is finite, i.e. no
ultraviolet (UV) cut off is necessary.

The IR divergency in the grand-canonical potential (for $T=0$ equal
 to the ground state
energy) led us to re-analyse the derivation of the CS theory.

\section{Formulation of the Chern--Simons Theory}
\subsection{Hamiltonian}\label{subsec:H}
We start with the Hamiltonian for the electrons on the
 physical lattice~\cite{WDK};
the lattice formulation allows
a rigorous implementation of the CS transformation.
\bea
\label{HLATT}
&&H
= -\frac{\hbar_{}^{2}}{2m_{b}^{}\alpha_{}^{2}}
    \sum \limits_{{\bf m} \atop j=1,\ldots,4 }
    \hat{\psi}_{}^{(e) \dagger}({\bf m})\,
    \left[
      U_{{\bf m}^{}\Leftarrow{\bf m}^{}+{\bf e}_{j}^{}}
      \hat{\psi}_{}^{(e)}({\bf m}+{\bf e}_{j}^{})
      -\hat{\psi}_{}^{(e)}({\bf m})
    \right]  \nonumber \\
&&+\frac{1}{2}
    \sum \limits_{{\bf m}_1, {\bf m}_2}
    \hat{\psi}_{}^{(e) \dagger}({\bf m}_{1}^{})
    \hat{\psi}_{}^{(e) \dagger}({\bf m}_{2}^{})\,
    \frac{\epsilon^2}
         {|{\bf r}_{{\bf m}_1}-{\bf r}_{{\bf m}_2} |}\,
    \hat{\psi}_{}^{(e)}({\bf m}_{2}^{})
    \hat{\psi}_{}^{(e)}({\bf m}_{1}^{})    \\
&& - \mbox{(Contribution from positive background)}
 -\mu
    \sum\limits_{{\bf m}^{}}
    \hat{\psi}_{}^{(e) \dagger}({\bf m})
    \hat{\psi}_{}^{(e)}({\bf m})  \; ,  \nonumber \\
\label{U}
& & U_{{\bf m}^{}\Leftarrow{\bf m}^{}+{\bf e}_{j}^{}}^{}
= \exp
  \left\{
    \frac{i|e|\alpha}{\hbar}\,
    A_{j}^{}({\bf m}^{}+ {\bf e}_j^{}/2) \nonumber
  \right\}     \; .
\eea
$\alpha$ is the lattice constant, $A_j({\bf m}+{\bf e}_j/2)$ the external
vector potential; ${\bf e}^{}_j$ are the vectors to the nearest neighbors on
the square lattice (${\bf m}$).
The electron field operator $\hat{\psi}^{(e)}$ satisfies
magnetic quasi-periodic boundary conditions.

The CS transformation relates the electron operator to the CF operator:
\be
\label{CS}
\hat{\psi}^{(e)}({\bf m}) =
\exp \left\{\frac{i|e|}{\hbar}\sum_{{\bf n}}
\left[
f^{(np)}({\bf m}) \rho _{el}
+ f^{(p)}({\bf m}-{\bf n}) (\hat{n}_{{\bf n}} - \rho _{el})
 \right]\right\}
\hat{\psi}({\bf m}) \; .
\ee
Here, $f^{(np)}$ is nonperiodic; it leads to periodic
boundary conditions for $\hat{\psi}$, and after insertion into
(\ref{HLATT}), it compensates the
external magnetic vector potential.
The $f^{(p)}$ is periodic, and the second term in the exponent corresponds to a
compensated CS flux.  We omit explicite formulae for
the $f$. The particle number for the electrons
or CF is $\hat{n}_{{\bf n}}$. Because the bases for the Hilbert
 spaces of $\hat{\psi}^{(e)}$
and $\hat{\psi}$ differ only by phase factors, the QM trace remains the
same.

Because we consider a diluted system ($l_B \gg \alpha $), we go with
(\ref{HLATT}) to the continuum approximation by expanding
the $\hat{\psi}({\bf m}+ {\bf e}_j^{})$ and the difference of the exponents
 from (\ref{CS}).
In the second order in the difference of the $f^{(p)}$, we obtain an interaction
Hamiltonian with six $\hat{\psi}$-operators in the order
$\hat{\psi}^{\dagger} \hat{n} \hat{n} \hat{\psi}$.
In the following, we retain this term's original form, which is
not normally ordered,
for further approximations by expansion in powers of the density
fluctuations.
Further, we refrain from approximating the external pair of operators,
$\hat{\psi}^{\dagger} \ldots \hat{\psi}$,
by the average electron density $\rho _{el}$.

\subsection{Path Integral Formulation}\label{subsec:PI}
We now formulate the theory by means of the coherent state Gra{\ss}mann
PI. For the imaginary time, we use a lattice with
a lattice constant $\Delta t \to 0 , \;
{\cal N} \Delta t = -i\hbar \beta$.
In order to avoid normal ordering of the
 $\hat{\psi}^{\dagger} \hat{n} \hat{n} \hat{\psi}$ term,
we use in the PI besides the time points $t_{\nu} , \; \nu = 0, 1, \ldots , {\cal N}-1$
also intermediate times, i.e. use $t_{\bar{\nu}} , \;
\bar{\nu} = 0, 1/2, 1, \ldots , {\cal N}-1/2$.
The partition function is
\bea
\label{PI}
Z&=&\int \prod\limits_{{\bf k}, \bar{\nu}}
\left\{ d\, \psi^{*}_{\bar{\nu}} ({\bf k})
d\, \psi^{}_{\bar{\nu}} ({\bf k}) \right\}
\exp \Big\{
        S_{}^{(0)}
        +S_{}^{({\cal A} {\cal A})}
        +S_{}^{(j{\cal A})} + S_{}^{(Coul)}
         \Big\}    \; ,  \\
\label{PI1}
S_{}^{(0)} =&&
 \sum\limits_{\bar{\nu} , {\bf k}}^{}
 \psi_{\bar{\nu}}^{*}({\bf k}) \left\{
  \psi_{\bar{\nu} - 1/2}({\bf k}) - \psi_{\bar{\nu}}({\bf k}) \right\}
   \nonumber  \\
 -&& \frac{i\Delta t}{\hbar}
   \sum\limits_{\nu , {\bf k}}^{}
   \{E^{0}({\bf k}) - \mu \} \psi_{\nu - 1/2}^{*}({\bf k})
   \psi_{\nu - 1}({\bf k})   \; , \nonumber \\
\label{PI2}
S_{}^{({\cal A} {\cal A})}
=
 &-&\frac{i\Delta t}{\hbar}
  \frac{\hbar ^2 \tilde{\phi}^2}{2m_b F^2}
  \sum\limits_{{\nu, {\bf k}_1}\atop{{\bf k}', {\bf k}''(\neq 0)}}
  \psi_{\nu}^{*}({\bf k}_1)
  \varrho _{\nu}({\bf k}')
  \varrho _{\nu - 1/2} ({\bf k}'')
  \psi_{\nu - 1}^{}({\bf k}_1 - {\bf k}' - {\bf k}'') \nonumber \\
  &&\ast \; \mbox{\boldmath$\kappa$}({\bf k}') \;
  \mbox{\boldmath$\kappa$}({\bf k}'')
  \; , \nonumber \\
\label{PI3}
S^{(j{\cal A})}
&=&
 -\frac{i\Delta t}{\hbar}
 \frac{\hbar^{2}\tilde \phi }{m_{b} F}
 \sum\limits_{{\nu, {\bf k}_1}\atop{{\bf k}(\neq 0)}}
 \psi_{\nu - 1/2}^{*}({\bf k}_1)
 {\bf k}_1
 \psi_{\nu - 1}({\bf k}_1  - {\bf k})
 \varrho _{\nu - 1/2} ({\bf k}) \;
  \mbox{\boldmath$\kappa$}({\bf k}) \; , \nonumber \\
\label{PI4}
S^{(Coul)}
&=&
 -\frac{i\Delta t}{\hbar}
 \frac{1}{2 F}
 \sum\limits_{{\nu, {\bf k}_1}\atop{{\bf k}(\neq 0)}}
 \psi_{\nu - 1/2}^{*}({\bf k}_1)
 \psi_{\nu - 1}({\bf k}_1 - {\bf k})
 \varrho _{\nu - 1/2} ({\bf k}) \;
 \frac{2\pi \epsilon^2}{k} \; , \nonumber \\
\label{PI5}
\varrho _{\nu}({\bf k})
&=&
 \sum_{{\bf k}_{1}}
 \psi_{\nu}^{*}({\bf k}_{1})
 \psi_{\nu - 1/2}({\bf k}_{1}^{}+{\bf k})   \; , \; \;
 \mbox{\boldmath$\kappa$}({\bf k}) = - 2 \pi i (e_z \times \frac{{\bf k}}{k^2})\;
 J_0(kR_0) \; . \nonumber
\eea
$J_0$ is the Bessel function,
$R_0$ a cut off ($\alpha \ll R_0 \ll l_B$) formally used in the continuum
approximation; for the diluted system distances between the CF of order $R_0$
are thermodynamically negligible.

Evaluation of the partition function (\ref{PI}) in low order perturbation
 theory
leads to the following results: \\
(1) The first order diagram for the grand-canonical potential
is now convergent. The absence of the IR divergency can be
 traced back to the order
of the operators $\hat{\psi}^{\dagger} \hat{n} \hat{n} \hat{\psi}$. Approximation of the
external $\hat{\psi}^{\dagger} \ldots \hat{\psi}$
by $\rho_{el}$ would lead to an UV divergency. \\
(2) Unlike the grand-canonical potential,  the CF self energy
shows an IR divergency in first order in the interaction:
\be
\Sigma^{(1)} (k) \sim \mp \ln \frac{k_F}{k_{min}} \;\;
\mbox{for} \; k < k_F \;\; \mbox{or} \;\; k > k_F \;, \; \mbox{respectively}.
\ee

The above considerations show that a PI approach, respecting the
 order of the operators
in the original Hamiltonian, provides a promising starting point
 for treating the IR divergencies in higher order
perturbation theory in the interaction.

\section*{Acknowledgments}
We like to acknowledge the financial support by the
Deutsche Forschungsgemeinschaft under grant nos. AP 47/1--2,
WE 480/2--2.

\end{document}